\documentclass[sigconf,nonacm]{acmart}
\usepackage{booktabs} 
\usepackage{lipsum}   
\usepackage{comment}
\usepackage{url}
\usepackage{hyperref}
\usepackage{amsmath,amsfonts}
\usepackage{caption}
\usepackage{longtable}
\usepackage{caption}
\usepackage{multirow}
\usepackage{booktabs}

\usepackage{tikz}
\usepackage{amsmath}
\usepackage{amsthm}
\theoremstyle{definition}

\usepackage{algorithmic}
\usepackage{graphicx}
\usepackage{textcomp}
\usepackage{xcolor}
\usepackage{filecontents}
\usepackage{algorithm}

\usepackage{listings}
\usepackage{xurl}
\definecolor{codegreen}{rgb}{0,0.6,0}
\definecolor{codegray}{rgb}{0.5,0.5,0.5}
\definecolor{codepurple}{rgb}{0.58,0,0.82}
\definecolor{backcolour}{rgb}{0.95,0.95,0.92}

\lstdefinestyle{mystyle}{
  backgroundcolor=\color{backcolour}, commentstyle=\color{codegreen},
  keywordstyle=\color{magenta},
  numberstyle=\tiny\color{codegray},
  stringstyle=\color{codepurple},
  basicstyle=\ttfamily\footnotesize,
  breakatwhitespace=false,         
  breaklines=true,                 
  captionpos=b,                    
  keepspaces=true,                 
  numbers=left,                    
  numbersep=5pt,                  
  showspaces=false,                
  showstringspaces=false,
  showtabs=false,                  
  tabsize=2
}

\title{Using Codebooks to Detect Cybercrime Topics in Text Narratives}

\author{Shufan Chai}
\authornote{The first 2 authors contributed equally.}
\affiliation{%
  \institution{Northeastern University}
  \city{Oakland}
  \state{CA}
  \country{USA}
}
\email{helloshufanfan@gmail.com}

\author{Liangliang Sun}
\authornotemark[1]
\affiliation{%
  \institution{Northeastern University}
  \city{Oakland}
  \state{CA}
  \country{USA}
}
\email{sun.liang@northeastern.edu}

\author{Jessica Staddon}
\affiliation{%
  \institution{Northeastern University}
  \city{Oakland}
  \state{CA}
  \country{USA}
}
\email{j.staddon@northeastern.edu}
\begin{abstract}
In the United States, management of cybercrime-related consumer complaints increasingly falls on state and city governments given de-staffing of federal agencies. AI, and in particular, large language models (LLMs), shows promise for detecting cybercrime in text  complaints, but often via specialized models that local governments are not resourced to develop and maintain. We present an LLM prompting method that uses  codebooks from qualitative cybercrime research to detect cybercrime topics in consumer narratives. For two cybercrime topics, impostor scams and identity theft, we demonstrate the method achieves high precision and recall across multiple runs of 5  models in the Gemini and GPT model families. This strategy suggests a path for resource-constrained organizations, like many local governments, to leverage frontier models to support community safety.

\end{abstract}

\begin{document}

\maketitle

\section{Introduction}

Ongoing resource reductions in federal consumer protection agencies in the United States (e.g., The Consumer Financial Protection Bureau \cite{cfpb, cfpb-layoffs}) are associated with US communities increasingly turning to state and city governments for help recovering from the effects of cybercrime, like scams and identity theft  (e.g. \cite{cho2024state,  widman2021protecting,  state_oversight25}). AI shows promise for detecting many forms of cybercrime, and local governments are increasingly deploying AI \cite{SF-copilot, miami-dade-ai}. However, many approaches to cybercrime detection involve customized models \cite{nicholas2024scamdetector, bitaab2025scamnet} that under-resourced organizations don't have the  training or data to support \cite{oecd2025governing, stern2025removing}. In addition, pre-trained models that can be more easily used, have been found to have deficiencies in security \cite{prakash2025learned,chen2023can, meiklejohn2026helpbench}.

In parallel, the cybercrime research community is conducting studies that deeply explore  cybercrime ecosystems.  
These complex and nuanced explorations often involve qualitative labeling of data (e.g., \cite{ricaldi2025uncovering, kelley2026generative}). Hence, a common side effect of research supporting cybercrime defense and remediation is human-designed documents that characterize members or  processes in cybercrime ecosystems through definitions, examples and counter-examples, also known as ``codebooks'' \cite{macqueen1998codebook}.

Inspired by the paradigm of using human-authored policies to guide pretrained LLMs to automate content moderation \cite{palla2025policy}, we ask whether the rich human-annotated data sets from cybercrime research can be paired with pre-trained LLMs for a generalist approach to the detection of cybercrime. More concretely we explore the following research question:

RQ: \textit{Can generalists use pretrained LLMs to accurately identify cybercrime in text narratives using researcher-authored codebooks?}

We explore this research question in the context of two prominent security problems, impostor scams and identity theft. Scams are a large and growing consumer safety risk, and impostor scams are reported by the FTC to be the most common form of scam \cite{ftc_top_scams, ftc-scam-report}. Impostor scams are characterized by the use of impersonation to enable scammers to benefit from pre-existing trust a target has in an institution (e.g., a bank) or a friend or family member, thus encouraging the target to take a financially harmful action. Identity theft is a financial harm in which a target's identity is appropriated to open new accounts in the person's name (E.g., credit card accounts) or gain access to existing accounts, without the target's permission.\footnote{While it is possible for an incident to involve both identity theft and an impostor scam, it is not required and Consumer Protection Agencies (e.g., the CFPB) treat them as separate risks.}

We iteratively developed codebooks for impostor scams and identity theft that support consistent and reliable labeling by humans. The authors of this paper used the codebooks to build a corpus of positive and negative examples of impostor scam narratives ($n=1357$) and a corpus of positive and negative identity theft examples ($n=981$) from the public Consumer Financial Protection Bureau (CFPB) complaints database. Across experiments with 5 LLMs from two leading model families (Gemini and GPT) we find that the average precision improvement is at least $.2$  when prompted with the codebooks ($.2$ for impostor scams and $.23$ for identity theft). In addition, all models but the optimized Gemini 2.5 Flash model, detect impostor scams and identity theft in CFPB narratives with more than $.8$ precision and recall, when prompted with the codebooks. The results suggest strong performance is possible while respecting budget-constraints when codebooks are used. The method is generalist-friendly as the prompt template does not require prompt engineering expertise.
In summary, we make the following contributions:
\begin{enumerate}
\item \textbf{Prompt Engineering:} A generalist method for detecting cybercrime incidents in text narratives using pretrained LLMs and researcher-authored codebooks.
\item \textbf{2 Cybercrime Case Studies:} Experimental evidence that the method improves precision over baseline performance and achieves high precision and recall with pretrained and nonoptimized models, for the cases of impostor scams and identity theft. The codebooks are available in this paper and the corpora are anonymously published (links in Appendix~\ref{app:openscience}).
\end{enumerate}

\subsection{Related Work}
\label{sec:relwork}
There is a long tradition of using AI in the areas of our use cases: scams and fraud \cite{chowdhury2024advancing}. Recently, LLMs have shown promise in automating the detection of scams in text narratives when fine-tuned to create specialized models (e.g., \cite{nicholas2024scamdetector, bitaab2025scamnet}). To support organizations that are not resourced to develop specialized models we study the effectiveness of techniques using pre-trained LLMs.

Authoritative content has been found to increase LLM accuracy when added to prompts in a number of domains includes areas of healthcare like psychological diagnosis \cite{sarma2025simulated}. This paradigm was first applied to automate user security and safety tasks by 
\cite{palla2025policy} and \cite{kholkar2025policy}. In
 \cite{palla2025policy}, a framework in which content moderation policies, originally developed for use by professional human moderators, are used in LLM prompts to partially automate content moderation, is presented. They provide an evaluation of the framework and an overview of the challenges to implementing the framework in an organization. Similarly, in \cite{kholkar2025policy}, the authors use software development content such as product design documents to inform an LLM-based approach to identifying and enforcing security guardrails. We extend \cite{palla2025policy,kholkar2025policy} by showing that human-authored codebooks can also be effective cybercrime detectors.

The study of cybercrime benefits from qualitative research to understand the ecosystem. For example, qualitative research has provided insights into scam defense \cite{legarda2026breaking,Harvey2026}, scammer techniques \cite{asyali2026fake, kelley2026generative, cole2024qualitative} and the target's experience of cybercrime \cite{cole2024qualitative}. These studies generally involve human-authored codebooks that could be re-used to automate detection via LLMs.

The use of LLMs for data annotation (or equivalently, topic and attribute prediction, as in this paper) was introduced in \cite{wang2021want} and has been used in a range of contexts since. However, most LLM-based annotation techniques that use codebooks or other qualitative research artifacts, do not directly apply codebooks, but instead use them to customize LLM prompts (e.g., \cite{he-etal-2024-annollm,xiao2023supporting, dunivin2025scaling}). In addition, these works often use smaller (e.g., open weight) models, and achieve reported performance that is well below our experiments, although sometimes better than baseline performance (\cite{halterman2026codebook, ruckdeschel2025just}). Prompt customization benefits both from LLM and domain expertise and so is not a generalist task.

Perhaps closest to this paper is \cite{relins2025using}, which uses human-authored codebooks to identify specific ``vulnerabilities'' (e.g., mental health problems) in police incident narratives. Using smaller models (Llama 8B, 70B and GPT-4o), \cite{relins2025using} reports precision that ranges from below $.2$ to at most $.7$. While this is weaker performance than in our experiments, it is compatible with our findings in that we find the optimized model, Gemini 2.5 Flash performs significantly worse than the larger models. We hypothesize that LLMs have only recently become capable of taking advantage of prompts with the complexity of human-authored codebooks.

Finally, we note that while this paper uses human-authored codebooks to automate annotation (and equivalently, prediction) via LLMs, recent research has developed methods for using LLMs to automate the creation of codebooks that can effectively be used by human annotators \cite{adeseye2026llm, zambrano2026data}.

\section{Data and Methodology}
\label{sec:data}

In this paper, we measure the accuracy of predictions made by an LLM when a codebook is and is not included in prompts. More formally, for a topic, $T$, prompt, $\mathcal{P}$, and narrative, $D$, an LLM, $\mathcal{L}$,  outputs $1$ if it predicts $D$ does discuss the topic, $T$, and outputs $0$ otherwise. Hence, we term a model and prompt a \textit{predictor}. In this paper we experiment with 2 topics, impostor scams and identity theft.

For simplicity of exposition, we often refer to the performance (e.g., precision and recall) of a predictor by referring to the prompt that defines it. In particular, the main comparison of this paper is between the performance of \textit{codebook prompts} and \textit{baseline prompts} (no codebook), each of which use the prompt template of Section~\ref{sec:prompts}.

 The following subsections describe the specific data sets and prompts in this paper.

\subsection{Data Sets}
\label{sec:data}
In Section~\ref{sec:case-studies}  we compare the performance of codebook-based prompts with the baseline performance of each of 5 models on data sets of positive and negative examples of impostor scams and identity theft. 
The data sets  of text narratives were built from a publicly available source,  complaints submitted to the Consumer Financial Protection Bureau (CFPB), a United States government agency responsible for consumer protection in the financial sector.
Since the CFPB was created  in 2011 through the Dodd-Frank Act\footnote{12 U.S. Code § 5491 - Establishment of the Bureau of Consumer Financial Protection}, the bureau has collected and monitored complaints from consumers related to financial safety in a variety of contexts (e.g., credit reporting, mortgage lending, automobile financing, student loans, scams and fraud). Consumers submitting complaints to the CFPB have the option of submitting a text narrative in addition to selecting from prepopulated options in several fields. If they submit a narrative and consent to making it public, it is redacted by the CFPB to reduce reidentification risk following the CFPB's ``scrubbing standard'' \cite{cfpb_personal}. We built 2 data sets from those in the CFPB database that include scrubbed narratives as described below. Both data sets are available at the link in Appendix~\ref{app:openscience}. 

\subsubsection{Impostor Scams Data Set}
\label{sec:impostordata}
While scams are a significant online safety risk, they are a small portion of the complaints received by the CFPB. To more efficiently build a set of positive examples (impostor scams) and negative examples (non-impostor scams or fraud) we used the ``issue'' selected by consumers who consented to including their (redacted) narratives in the CFPB database as a filter. The issues available to complainants that are most related to scams are  ``fraud or scam'' and ``Problem with fraud alerts or security freezes''.

The definitional distinction between scams and non-scam fraud is that in 
scams, the user (or, complainant) takes self-harming actions \cite{modic2013scam}. In particular, while both types of
complaints may involve transactions considered fraudulent, in the case of a scam a user is tricked
into authorizing the transaction themselves, whereas in non-scam fraud a bad actor authorizes the
transaction (e.g., using stolen credentials). That said, the terms ``fraud'' and ``scam'' are often used interchangeably, and so we started with the $34,015$ public narratives labeled with either the ``Fraud or scam'' or ``Problem with fraud alerts or security freezes'' issue that were available on April 18, 2025 in the CFPB Complaints Database to build a data set of impostor scams and non-impostor fraud or scams.

Following the common qualitative research practice of using a codebook for data labeling \cite{macqueen1998codebook}, each of the 3 authors first independently labeled the same set of 100 narratives ``impostor scam'' or ``not impostor scam'', sampled from the $34,015$ public narratives, using an initial codebook consisting of a brief definition. The authors met to discuss cases of disagreement, clarify label definitions, and resolve disagreements. Based on the discussion, we updated and refined the codebook to expand on the definition and include positive and negative examples.
 
To assess whether all three authors had a shared understanding of the refined codebook, we independently labeled an additional shared set of 10 narratives and computed pairwise inter-rater agreement using Cohen's kappa \cite{cohen1960coefficient,gisev2013interrater}. The resulting pairwise kappas (0.80, 1.00, 0.80) indicated strong agreement across all authors and a consistent interpretation of the codebook definitions.

The complete final codebook is in Table~\ref{tab:impostor-scam-codebook} and an excerpt  is in Table~\ref{tab:impostor-codebook-partial}.
 
After achieving high inter-rater agreement, each author coded disjoint subsets of the data, resulting in a data set of $1357$ narratives after duplicate narratives were removed. During experiments in support of the method of this paper, labeled samples were reviewed by all authors and  $38$ labels were adjusted. The final labeled data set of $1357$ consists of $187$ impostor scams ($13.8\%$ of the data set). This data set is denoted as $\mathcal{D}_{IS}$ in the rest of this paper.

\begin{table*}[htbp]
    \centering
    \fbox{%
        \parbox{0.95\textwidth}{
            \centering
            \begin{tabular}{l}
               A scammer poses as a member of a trusted organization (e.g., bank or credit union, a government department, the police)\\ or having an authoritative role (e.g., IRS agent, fraud specialist) or as a friend or family member to use a pre-existing trust\\ relationship to facilitate a scam.\\
\\
Definition scope\\
\\
What is not an imposter scam?\\
  \hspace*{0.5cm}Romance scams – not using authority or impersonation, gain trust in other ways\\
  \hspace*{0.5cm}More traditional confidence scams\\
  \hspace*{0.5cm}Puppy scams are not imposter scams, unless posing as a well-known breeder\\
  \hspace*{0.75cm}  Fake products/services (unless impersonating a reputable goods or services provider)\\
  \hspace*{0.5cm}If posing but not making use of authority to persuade, e.g.\\
   \hspace*{0.75cm}``I was a victim of a fraudulent scheme in XX/XX/year> that involved Bank XXXX XXXX XXXX product.\ldots''\\
  \hspace*{0.5cm} \ldots\\\\

What is an imposter scam?\\
  \hspace*{0.5cm}Fake rental scams (even when vague): there is evidence the ``landlord'' didn’t actually have the property\\
  \hspace*{0.5cm}this does not include claimed scams in which rent payments are disputed\\
  \hspace*{0.5cm}In absence of clear evidence for or against an imposter scam we assume a rental scam is an imposter scam\\
 \hspace*{0.5cm}Fake fraud alert scams\\
 \hspace*{0.5cm} \ldots\\
 
            \end{tabular}
        }%
    }
    \caption{An excerpt from the impostor scam codebook showing the structure and most of the content. The complete codebook is in Appendix~\ref{app:codebooks}. This codebook was iteratively developed by the authors of this paper using the process described in Section~\ref{sec:impostordata}.}
    \label{tab:impostor-codebook-partial}
\end{table*}

\subsubsection{Identity Theft Data Set}
\label{sec:identitydata}
While the term ``identity theft'' is often used broadly to refer to financial harms \cite{hoofnagle2007identity}, in this paper we follow the definition given by Congress in United States Code Title 18, Section 1028 \cite{uscode18-1028}; in short, the use of an individual's legal identification information to assume their identity for the purposes of creating or taking over new accounts and causing financial harm.

Although identity theft continues to be a common financial harm  \cite{ftc2024sentinel}, it is, like impostor scams, a relatively small portion of the CFPB Complaints database, so we  used a sampling strategy to identify positive examples of identity theft. The densest proportion of identity theft narratives appear to be in the 1,722 CFPB complaints for which complainants selected the label ``Identity theft / Fraud / Embezzlement'', however, that label was deprecated in late 2017 and so we augmented our initial sample with narratives from product areas such as credit cards, which are often involved in identity theft. Our starting data pool had approximately 2000 narratives. As when building the impostor scam data set,  each of the 3 authors first independently labeled the same set of 100 narratives ``identity theft'' or ``not identity theft'' using an initial codebook consisting of a brief definition. The authors met to discuss cases of disagreement, clarify label definitions, and resolve disagreements. Based on the discussion, we updated and refined the codebook to expand on the definition and include positive and negative examples.

The complete identity theft codebook is in Table~\ref{tab:identity-theft-codebook} and an excerpt is in Table~\ref{tab:identity-codebook-partial}.
 
To assess whether all three authors had a shared understanding of the refined codebook, we independently labeled an additional shared set of 10 narratives and achieved perfect agreement across all authors. Finally, each author coded disjoint subsets of the data, and duplicate narratives were removed, resulting in a data set of $981$ narratives, consisting of $427$ narratives describing identity theft ($43.5\%$ of the data set). This data set is denoted as $\mathcal{D}_{IT}$ in the rest of this paper.

\begin{table*}[htbp]
    \centering
    \fbox{%
        \parbox{0.95\textwidth}{
            \centering
            \begin{tabular}{l}
              Identity theft is fraud committed or attempted by using the identifying information of another person without his or her \\authority. Identifying information may include such things as a Social Security number, account number, date of birth, driver's \\license number, passport number, biometric data and other unique electronic identification numbers or codes.\\
\\
You may be a victim of identity theft if you:\\
 \hspace*{0.5cm}Receive credit cards that you did not apply for.\\
 \hspace*{0.5cm}Receive bills or collection letters from companies that you never heard of or for accounts you don’t recognize.\\
 \hspace*{0.5cm}Receive rejection letters for loans you did not apply for.\\
  \hspace*{0.5cm}\ldots\\\\

What isn’t identity theft?\\
 \hspace*{0.5cm}Unauthorized use of credit cards is unlikely to involve identity theft since credit card transactions can often complete without\\ 
 \hspace*{0.75cm}without the need to assume the card owner’s identity\\
 \hspace*{0.5cm}Unauthorized credit card use is more commonly termed “credit card fraud” under federal law (15 U.S.C. § 1644 or \\
 \hspace*{0.75cm}18 U.S.C. § 1029)\\
 \hspace*{0.5cm}If the only clear financial harm is fraudulent CC charges, we don’t mark it as identity theft\\
 \hspace*{0.5cm}A breach or exposure of identity information can lead to identity theft, but is not on its own identity theft\\
 \hspace*{0.5cm}Unauthorized activation of a credit card; activation may require partial identity information (e.g. last 4 digits of an SSN), \\
 \hspace*{0.75cm}but not enough information to appropriate identity in a sustained manner\\
 \hspace*{0.5cm}If a scammer tells a target that they have a debt the target didn’t authorize, we do not consider this identity theft unless the \\
 \hspace*{0.75cm}narrative provides evidence that the debt truly exists. It is common for scammers to try to scare targets into believing they \\
 \hspace*{0.75cm}have an imaginary debt so we require evidence of the debt to label it identity theft.\\
 \hspace*{0.5cm}A credit check may require identity information but it isn’t an appropriation of the target’s identity and so it is not identity \\
 \hspace*{0.5cm}theft\\
 \\
What is identity theft?\\
\hspace*{0.5cm}If the narrative states that identity theft has occurred and there is no evidence in the remainder of the narrative indicating\\
\hspace*{0.75cm}otherwise, we mark it as identity theft (even if there is no additional supporting evidence beyond the identity theft \\
\hspace*{0.75cm}assertion).\\
\hspace*{0.5cm}If the main complaint of a narrative are events happening after identity theft, and not about the identity theft itself, we still \\
\hspace*{0.75cm}mark the narrative as identity theft since the goal is to identify complaints related to cases of identity theft\\
\hspace*{0.5cm}A process that involves identity verification (e.g., mail forwarding) is identity theft\\
\hspace*{0.5cm}If new financial accounts are associated with a complainant without authorization (e.g., they appear on a credit report), we \\
\hspace*{0.75cm}assume identity theft has occurred since account creation requires identity information.\\
\hspace*{0.5cm}If a complaint requests the debt not be reported to credit agencies we assume this is because the debt has been reported and\\ 
\hspace*{0.75cm}so constitutes identity theft\\
\hspace*{0.5cm}\dots
            \end{tabular}
        }%
    }
    \caption{An excerpt from the identity theft codebook showing the structure and almost all of the content. The complete codebook is in Appendix~\ref{app:codebooks}.This codebook was iteratively developed by the authors using the process of Section~\ref{sec:identitydata}. Note that the portion of the codebook that begins with ``You may be a victim\ldots'' and ends with ``\ldots in your name.'', is taken from public information from the Texas Attorney General website \cite{texas_ag_identity_theft}.}
    \label{tab:identity-codebook-partial}
\end{table*}

\subsection{Prompts}
\label{sec:prompts}
The prompt engineering strategy of this paper incorporates codebooks in the following simple template. We use this template for both the impostor scam and identity theft experiments. In the former, the second sentence of the prompt is ``Your task is to determine whether a consumer complaint describes an **Imposter Scam**''\footnote{Both ``imposter'' and ``impostor'' are common spellings and we did not observe any LLM sensitivity to the spelling choice.}, and in the latter the second sentence is ``Your task is to determine whether a consumer complaint describes **Identity Theft**''.

\begin{lstlisting}[language=Python]
LLM_instruction = ("You are a professional fraud detection analyst. Your task is to determine whether a consumer complaint describes {an **Imposter Scam**, **Identity Theft**}."
 {Codebook}
 "Determine whether the following consumer complaint describes  {an imposter scam, identity theft}. Answer only 'yes' or 'no' without explanations."
{Consumer complaint: {narrative}})
\end{lstlisting}

This template does not assume substantial prompt engineering expertise. Indeed the only aspects of prompt engineering best practices present in the template that recent studies have found to \textit{not} be commonly used \cite{jin2025understanding} are a role (``professional fraud detection analyst'') and output formatting (``Answer only `yes' or `no'...'').

\begin{figure*}
    \centering
    \includegraphics[width=\linewidth]{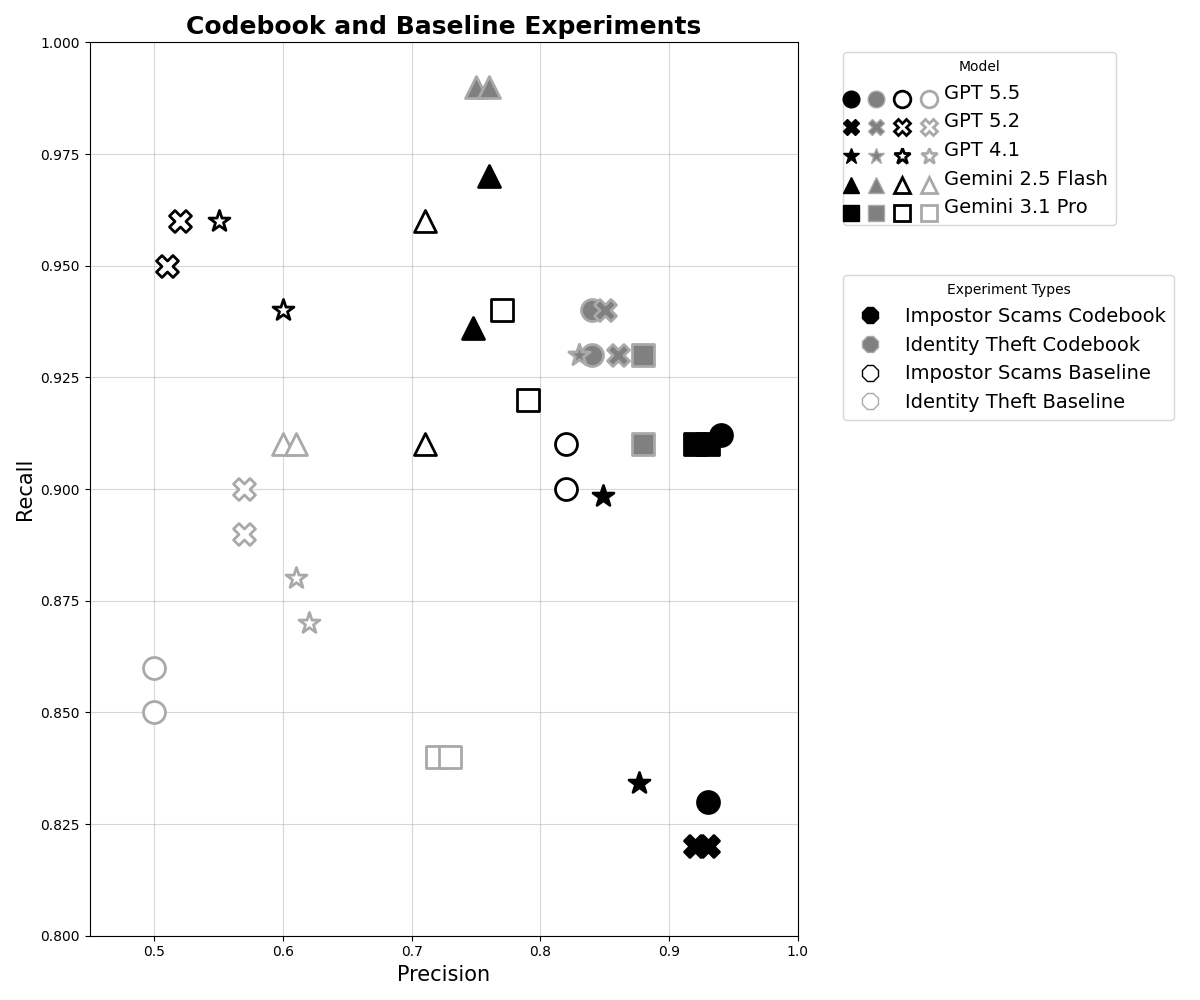}
    \caption{The precision and recall measurements for the 20 impostor scam experiments and 20 identity theft experiments described in Section~\ref{sec:case-studies}. For all models and both cybersecurity topics, the precision of the codebook prompt is greater than the baseline precision. The precision improvement is generally larger for identity theft, for which the baseline prompt performance is poor.}
    \label{fig:codebook-performance}
\end{figure*}

In this paper, we refer to an instantiation of the prompt template with the name of the codebook, i.e., \textit{impostor scams codebook} or \textit{identity theft codebook}. If no codebook is inserted, we term the prompt \textit{baseline}.

\section{Case Studies}
\label{sec:case-studies}
For each codebook, we ran the codebook prompt with each of the 5 models twice. Similarly, we ran the baseline prompt twice with each of the five models. Performance of prompts was measured using precision and recall \cite{manning2008introduction}. Experiments with Gemini 2.5 Flash were conducted in November-December 2025 for impostor scams and April-May for identity theft. Experiments with GPT 4.1 were
run in January 2026 for impostor scams and April-May for identity theft. Experiments with GPT 5.2 and 5.5 and with
Gemini 3.1 Pro, were run in April and May of 2026. In keeping
with the goal of developing a method appropriate for generalists,
experiments were run with the default parameters when possible.

\subsection{Results}
\label{sec:results}
All models substantially improve in precision when codebooks are added to the prompt and with the exception of the Gemini 2.5 Flash model, all models exceed $.8$ precision with codebooks. The improvement is particularly large with identity theft, for which baseline performance is weak. We describe the results in more detail in the following.\\

\noindent
\textit{Impostor Scams.} The baseline performance of the models varies considerably for the 5 LLMs. The most recent models in each family, GPT 5.5 and Gemini 3.1 Pro, both meet or exceed a precision of $0.8$, whereas GPT 4.1 and 5.2 perform only moderately better than random, and Gemini 2.5 Flash has a precision of .71. Precision increases for the codebook prompt with \textit{all} models; the minimum is $.75$ (Gemini 2.5 Flash) and the maximum is $.94$ (GPT 5.5). The average across all codebook runs of all models (10 experiments) is $.88$ precision and $.88$ recall. The average precision improvement between the codebook and baseline prompts is $.2$.\\

\noindent
\textit{Identity Theft.} The baseline identity theft performance across models is consistently poor; precision ranges from $.5$ to $.73$ with an average of $.6$. As with impostor scams, the codebook prompt performs better across \textit{all} models. The minimum precision of the codebook prompt is $.75$ (Gemini 2.5 Flash) and the maximum precision is $.88$ (Gemini 3.1 Pro); the average across all runs of all models (10 experiments) is $.83$ precision and $.94$ recall. The average precision improvement between the codebook and baseline prompts is $.23$.

Figure~\ref{fig:codebook-performance} shows the combined precision and recall results and the complete numerical results are in Table~\ref{tab:impostor_scams_metrics} of Appendix~\ref{app:codebooks}.

\subsection{Discussion}

While most consumers are likely familiar with the terms ``impostor scam'' and ``identity theft'', the baseline results suggests that the ``understanding'' the LLMs in our experiments have of identity theft is not well-aligned with the codebook's definition (Table~\ref{tab:identity-theft-codebook}). This is compatible with the fact that publicly available Consumer Protection Agency (CPA) information about identity theft, while not at odds with the codebook, does not clearly reinforce it. For example, guidance from the Federal Trade Commission\footnote{\url{https://consumer.ftc.gov/articles/what-know-about-identity-theft}} (FTC) and Federal Deposit Insurance Corporation\footnote{\url{https://www.fdic.gov/consumer-resource-center/cybersecurity}} (FDIC) allows for unauthorized purchases with existing credit cards in their broad descriptions of identity theft. In contrast, following federal law, the codebook only considers \textit{new} credit cards to be within scope, since opening new accounts requires the theft of identity information.  This ambiguity in the definition of identity theft is likely well-represented in model training data, hence the poor baseline performance. In contrast, CPAs consistently align with the codebook in characterizing impostor scams as scams in which a trusted person or organization is impersonated, and impostor scams has a far stronger baseline performance.

It is noteworthy that despite the likely poor alignment between model training data related to identity theft and the codebook, there is a similar average precision improvement with the codebook prompt ($.23$ for identity theft vs $.2$ for impostor scams), and, with the exception of the optimized Gemini 2.5 Flash model, all models exceed $.8$ precision with the identity theft codebook prompt. 

While these case studies suggest codebooks can form the basis of accurate LLM-based predictors, there are risks and limitations that we discuss below.

\textit{Low Quality or Inappropriate Codebooks.} This method assumes that codebooks have been developed so that independent human annotators can use them to reliably and consistently to annotate data \cite{macqueen1998codebook}. Indeed, sycophantic behavior of models, highlights the importance of trustworthy codebooks \cite{sharma2024towards}. Only codebooks that include evidence of rigorous development should be used. 

Perhaps a more significant concern is a mismatch between the codebook's intent and its use as a predictor. For example, the identity theft codebook (Table \ref{tab:identity-theft-codebook}) accepts a consumer's claim of identity theft provided there is no clear evidence to the contrary. A conservative choice like this may not be appropriate for some predictor use cases. In order to make the method appropriate for generalists, shared codebooks should come with a generalist-appropriate description of the codebook's goal and coverage. The Open Science community provides guidance for linking the provenance and data generation methodology with the data itself (e.g., via ``datasheets'', \cite{gebru2021datasheets,wilkinson2016fair}) and could be extended with sharing guidance tailored to codebooks to support the LLM predictor use case.

In addition, as noted in \cite{neumann2026not}, any LLM-based safety strategy needs to be supported by a governance framework.

\textit{Over-confident LLMs.}
It is possible that a ``confident'' model with an understanding that is at odds with a codebook will ignore the codebook. We do not see evidence of this in our case studies; precision of all models improves with codebook prompts. That said, incorporating prompt techniques for eliciting confidence
and/or reasoning, to force the predictor to explain how predictions
are compatible with the codebook may dimish this risk [37, 39].

\textit{Generalizability.} The case studies demonstrate this method shows promise for building predictors of cybercrime attributes or events based on text input, that lend themselves to qualitative analysis, and hence, the development of codebooks. While this scope of generalizability is fairly narrow, it is worth considering as AI capabilities expand, whether other cybercrime research artifacts can be repurposed to enable customized AI solutions.

\textit{Abuse.} When a detection method is made public there is a risk that future inputs could
be modified to game the outcome of the method; for example, historically, spammers used spam filter knowledge to modify their
email messages to avoid spam classification and reach more
inboxes \cite{rao2012economics}. With access to a codebook, a consumer could craft a narrative that is more likely to be detected as a desired form of cybercrime. If a codebook contains objectively verifiable attributes, this can help more quickly identify narratives that misled the detector.

\section{Conclusion and Open Problems}
We have presented a generalist-friendly method for building cybercrime predictors using codebooks and showed that it can be used to accurately predict impostor scams and identity theft, even when the baseline performance  of models (without a codebook), is poor. The method is appropriate for resource-constrained organizations in that it performs well with a variety of pretrained models (including those that may be cheaper to use) and does not require domain expertise. 

While these case studies are reason to be optimistic about this method, more experimentation is needed to understand how broadly the method can used within the cybercrime domain. In particular, in addition to testing with more topics or other cybercrime attributes, we have only experimented with a single prompt template and a more complex template that requires the model to justify predictions using the codebook, may further improve performance. It would also be useful to experiment with more codebooks, of varying complexity, to better understand the limits of what state of the art models can leverage and manage abuse risk. We also note that frontier model companies and annotation companies routinely develop evaluation rubrics that human reviewers use to gauge appropriateness of model behavior (e.g., \cite{phuong2024evaluating}) and these rubrics may also form the basis of useful prompts.

Finally, we note that this short paper does not address how a detector using the method of this paper can be safely integrated into a governance framework. In particular, as \cite{neumann2026not} observes for the content moderation context, use of LLM-based detectors should be in combination with other techniques (e.g., community-based moderation) that may more rapidly respond to changes in bad actor strategy.

\bibliographystyle{ACM-Reference-Format}
\bibliography{references,sample} 

\appendix 

\section{Open Science} 
\label{app:openscience}

We have anonymously published our impostor scam and identity theft data sets at the following URL: \url{https://docs.google.com/spreadsheets/d/e/2PACX-1vRqRy7c5gNEiD-i-tNtWl8_XuJeP_9BazPQqRa1F7lOBJHX6SA6iZ_2Rke8QLby-A/pubhtml}

The human-authored codebooks are in Section~\ref{sec:impostordata} (impostor scams) and Section~\ref{sec:identitydata} (identity theft).

The precision and recall calculations use well-known formulae \cite{manning2008introduction} and were implemented both in spreadsheets and notebooks.

\section{Ethical Considerations} 

An additional ethical consideration is the data we use, the complaints data from the Consumer Financial Protection Bureau (CFPB) complaints database \cite{cfpb}. The database is publicly available and released through the CFPB’s Consumer Complaint Database \cite{cfpb}. Complaints are published only if the consumer opts in to share their narrative publicly at the time of submission, and consumers may withdraw this consent at any time. In addition, before narratives are published, the CFPB applies a narrative scrubbing process to remove personal information that could directly identify an individual \cite{cfpb_personal}. This process includes automated checks and human review to try to ensure that personal identifiers are not present in the publicly released data. Consumers are informed about the consent process and the CFPB’s review procedures before they choose to publish their narrative.

\section{Use of Generative AI}
The topic of this paper is LLM-enabled identification of cybercrim in text naratives and we have have experimented with LLMs to develop and evaluate the proposed approach.

The prose of this paper was human-authored. We have occasionally used generative AI (primarily, Gemini and Claude) as an assistant in the following ways:
\begin {itemize}
\item To generate or debug code snippets for analysis or figure generation
\item To generate latex for tables and bibtex entries for references
\item To critique earlier paper drafts for clarity.
\end{itemize}

\section{Experiments and Codebooks}
\label{app:codebooks}

\begin{table*}[htbp]
\centering

\resizebox{\textwidth}{!}{%
\begin{tabular}{lcccccccc}
\toprule
Model & \multicolumn{2}{c}{\textit{Impostor Scam Baseline}} & \multicolumn{2}{c}{\textit{Impostor Scam Codebook}} & \multicolumn{2}{c}{\textit{Identity Theft Baseline}} & \multicolumn{2}{c}{\textit{Identity Theft Codebook}}  \\
\cmidrule(lr){2-3} \cmidrule(lr){4-5} \cmidrule(lr){6-7} \cmidrule(lr){8-9} 
 & Precision & Recall & Precision & Recall & Precision & Recall & Precision & Recall \\ 
\midrule
\multirow{2}{*}{GPT 5.5} & 0.82 & 0.90 & 0.94 & 0.91 & 0.50 & 0.86 & 0.84 & 0.94 \\
& 0.82 & 0.91 & 0.93 & 0.83 & 0.50 & 0.85 & 0.84 & 0.93 \\
 
\midrule
\multirow{2}{*}{GPT 5.2}& 0.52 & 0.96 & 0.92 & 0.82 & 0.57 & 0.90 &  0.86 & 0.93 \\
 & 0.51 & 0.95 & 0.93 & 0.82 & 0.51 & 0.95 & 0.85 & 0.94\\ 
\midrule
\multirow{2}{*}{GPT 4.1} & 0.60 & 0.94 & 0.88 & 0.83 & 0.61 & 0.88 & 0.83 & 0.93  \\
 & 0.55 & 0.96 & 0.85 & 0.90 & 0.62 & 0.87 & 0.83 & 0.93 \\
 
\midrule
\multirow{2}{*}{Gemini 2.5 Flash} & 0.71 & 0.91 & 0.76 & 0.97 & 0.6 & 0.91 & 0.75 & 0.99  \\
 & 0.71 & 0.96 & 0.75 & 0.94 & 0.61 & 0.91 & 0.76 & 0.99 \\
\midrule
\multirow{2}{*}{Gemini 3.1 Pro} & 0.79 & 0.92 & 0.92 & 0.91 & 0.72 & 0.84 & 0.88 & 0.91 \\
 & 0.77 & 0.94 & 0.93 & 0.91 & 0.73 & 0.84 & 0.88 & 0.93 \\
\bottomrule
\end{tabular}%
}
\caption{Precision and recall of baseline and codebook experiments for impostor scams and identity theft and all LLMs.}
\label{tab:impostor_scams_metrics}
\end{table*}

\begin{table*}[htbp]
    \centering
    \fbox{%
        \parbox{0.95\textwidth}{
            \centering
            \begin{tabular}{l}
               A scammer poses as a member of a trusted organization (e.g., bank or credit union, a government department, the police)\\ or having an authoritative role (e.g., IRS agent, fraud specialist) or as a friend or family member to use a pre-existing trust\\ relationship to facilitate a scam.\\
\\
Definition scope\\
\\
What is not an imposter scam?\\
  \hspace*{0.5cm}Romance scams – not using authority or impersonation, gain trust in other ways\\
  \hspace*{0.5cm}More traditional confidence scams\\
  \hspace*{0.5cm}Puppy scams are not imposter scams, unless posing as a well-known breeder\\
  \hspace*{0.75cm}  Fake products/services (unless impersonating a reputable goods or services provider)\\
  \hspace*{0.5cm}If posing but not making use of authority to persuade, e.g.\\
   \hspace*{0.75cm}``I was a victim of a fraudulent scheme in XX/XX/year> that involved Bank XXXX XXXX XXXX product. The perpetrators\\
  \hspace*{0.75cm}posed on XXXX as a nearby homeowner desiring to give away a grand piano. When I inquired, the response was the \\
  \hspace*{0.75cm}homeowner had given up and the piano had been moved. The scammer was willing to have me pay to have the piano\\
  \hspace*{0.75cm}moved back to Massachusetts. Suspicious, I called the BOA fraud line, and the operator looked up the moving company \\
  \hspace*{0.75cm}online, as I had, and assured me it look ok. Using XXXX, I paid {\$XXXX} for the move. A second communication\\
  \hspace*{0.75cm}indicated the piano would need to be packed more securely for a move, and more money was necessary. I no longer have\\
  \hspace*{0.75cm}a record of what I paid for that, through XXXX, but it was around {\$XXXX}. Needless to say, no piano was ever delivered.\\
  \hspace*{0.75cm}I contacted BOA, seeking fraud protection or reimbursement, and that was denied by the bank. I did not pursue the matter\\
  \hspace*{0.75cm}further.''\\
  \hspace*{0.5cm}A scammer uses a target’s personal information to impersonate them at financial institutions. This is identity theft but not an\\ \hspace*{0.5cm}impersonation scam unless the scammer acquires the personal information through an impersonation scam.\\
  \hspace*{0.5cm}Clear expression of imposter behavior is required\\
  \hspace*{0.5cm}Not enough to just include imposter-related terms like ``pose'' or ``clone''\\\\

What is an imposter scam?\\
  \hspace*{0.5cm}Fake rental scams (even when vague): there is evidence the ``landlord'' didn’t actually have the property\\
  \hspace*{0.5cm}this does not include claimed scams in which rent payments are disputed\\
  \hspace*{0.5cm}In absence of clear evidence for or against an imposter scam we assume a rental scam is an imposter scam\\
 \hspace*{0.5cm}Fake fraud alert scams\\
 \hspace*{0.5cm}Fake relative in peril – even if don’t explicitly say communicated with the impersonated relative. (e.g., ``On XX/XX/year> I\\
 \hspace*{0.5cm}received a phone call from a person claiming that my daughter was in jail and I needed to send money for her release. I made\\
 \hspace*{0.5cm}two separate transactions, through CashApp for {\$760.00} and {\$320.00} and then realized it was a scam. I tried to reach the\\
 \hspace*{0.5cm}company and stop the payment but only got a recording. I followed the steps to report it but received no response. I even\\ 
 \hspace*{0.5cm}requested a refund from the person 's CashApp only for it to be declined.'')\\
 \hspace*{0.5cm}Fake job scams: If someone poses as a representative of, or recruiter for, a seemingly reputable company\\
 \hspace*{0.5cm}Fake charity: someone poses as a representative of any charity (not necessarily well-known)
            \end{tabular}
        }%
    }
    \caption{The complete impostor scam codebook. This codebook was iteratively developed by the authors of this paper using the process described in Section~\ref{sec:impostordata}.}
    \label{tab:impostor-scam-codebook}
\end{table*}

\begin{table*}[htbp]
    \centering
    \fbox{%
        \parbox{0.95\textwidth}{
            \centering
            \begin{tabular}{l}
              Identity theft is fraud committed or attempted by using the identifying information of another person without his or her \\authority. Identifying information may include such things as a Social Security number, account number, date of birth, driver's \\license number, passport number, biometric data and other unique electronic identification numbers or codes.\\
\\
You may be a victim of identity theft if you:\\
 \hspace*{0.5cm}Receive credit cards that you did not apply for.\\
 \hspace*{0.5cm}Receive bills or collection letters from companies that you never heard of or for accounts you don’t recognize.\\
 \hspace*{0.5cm}Receive rejection letters for loans you did not apply for.\\
  \hspace*{0.5cm}Receive notices reflecting that you traveled to, lived in or did business in a jurisdiction to which you have no connections.\\
  \hspace*{0.5cm}Get calls from debt collectors or businesses about merchandise or services you did not buy.\\
  \hspace*{0.5cm}Fail to receive your bills or regular mail. (The ID thief may have changed your billing address!)\\
  \hspace*{0.5cm}Receive unexpected notices from the IRS about failing to report all your income or informing you that they received more\\ 
  \hspace*{0.5cm}than one income tax return in your name.\\
\\

What isn’t identity theft?\\
 \hspace*{0.5cm}Unauthorized use of credit cards is unlikely to involve identity theft since credit card transactions can often complete without\\ 
 \hspace*{0.75cm}without the need to assume the card owner’s identity\\
 \hspace*{0.5cm}Unauthorized credit card use is more commonly termed “credit card fraud” under federal law (15 U.S.C. § 1644 or \\
 \hspace*{0.75cm}18 U.S.C. § 1029)\\
 \hspace*{0.5cm}If the only clear financial harm is fraudulent CC charges, we don’t mark it as identity theft\\
 \hspace*{0.5cm}A breach or exposure of identity information can lead to identity theft, but is not on its own identity theft\\
 \hspace*{0.5cm}Unauthorized activation of a credit card; activation may require partial identity information (e.g. last 4 digits of an SSN), \\
 \hspace*{0.75cm}but not enough information to appropriate identity in a sustained manner\\
 \hspace*{0.5cm}If a scammer tells a target that they have a debt the target didn’t authorize, we do not consider this identity theft unless the \\
 \hspace*{0.75cm}narrative provides evidence that the debt truly exists. It is common for scammers to try to scare targets into believing they \\
 \hspace*{0.75cm}have an imaginary debt so we require evidence of the debt to label it identity theft.\\
 \hspace*{0.5cm}A credit check may require identity information but it isn’t an appropriation of the target’s identity and so it is not identity \\
 \hspace*{0.5cm}theft\\
 \\
What is identity theft?\\
\hspace*{0.5cm}If the narrative states that identity theft has occurred and there is no evidence in the remainder of the narrative indicating\\
\hspace*{0.75cm}otherwise, we mark it as identity theft (even if there is no additional supporting evidence beyond the identity theft \\
\hspace*{0.75cm}assertion).\\
\hspace*{0.5cm}If the main complaint of a narrative are events happening after identity theft, and not about the identity theft itself, we still \\
\hspace*{0.75cm}mark the narrative as identity theft since the goal is to identify complaints related to cases of identity theft\\
\hspace*{0.5cm}A process that involves identity verification (e.g., mail forwarding) is identity theft\\
\hspace*{0.5cm}If new financial accounts are associated with a complainant without authorization (e.g., they appear on a credit report), we \\
\hspace*{0.75cm}assume identity theft has occurred since account creation requires identity information.\\
\hspace*{0.5cm}If a complaint requests the debt not be reported to credit agencies we assume this is because the debt has been reported and\\ 
\hspace*{0.75cm}so constitutes identity theft\\
\hspace*{0.5cm}If a financial institution, or an employee of a financial institution, opens an account without authorization it is identity theft\\
\hspace*{0.5cm}If a financial institution, or an employee of a financial institution, runs a credit inquiry without authorization (since doing so \\
\hspace*{0.75cm}requires identity information)
            \end{tabular}
        }%
    }
    \caption{The complete identity theft codebook. This codebook was iteratively developed by the authors using the process of Section~\ref{sec:identitydata}. Note that the portion of the codebook that begins with ``You may be a victim\ldots'' and ends with ``\ldots in your name.'', is taken from public information from the Texas Attorney General website \cite{texas_ag_identity_theft}.}
    \label{tab:identity-theft-codebook}
\end{table*}

\end{document}